\documentclass[conference]{IEEEtran}
\IEEEoverridecommandlockouts
\usepackage[letterpaper, left=1in, right=1in, bottom=1in, top=0.75in]{geometry}

\usepackage{cite}
\usepackage{amsmath,amssymb,amsfonts}
\usepackage{algorithmic}
\usepackage{graphicx}
\usepackage{subfigure}
\usepackage{textcomp}
\usepackage{acronym}
\usepackage{multirow}

\usepackage{prettyref}
\newrefformat{fig}{Fig.~\ref{#1}}
\newrefformat{conj}{Conjecture~\ref{#1}}
\newrefformat{tab}{Table~\ref{#1}}
\newrefformat{sec}{Section~\ref{#1}}
\newrefformat{subsec}{Section~\ref{#1}}
\newrefformat{subsubsec}{Section~\ref{#1}}
\usepackage[super]{nth}
\usepackage[binary-units]{siunitx}
\usepackage{amsthm}
\usepackage{bm}

\usepackage[right=1in]{geometry}
 
\theoremstyle{definition}

\theoremstyle{remark}

\begin{document}

\title{Selection of Waveform Parameters\\
Using Machine Learning for 5G and Beyond
\thanks{\textcopyright 2019 IEEE.  Personal use of this material is permitted.  Permission from IEEE must be obtained for all other uses, in any current or future media, including reprinting/republishing this material for advertising or promotional purposes, creating new collective works, for resale or redistribution to servers or lists, or reuse of any copyrighted component of this work in other works.}}

\author{
\IEEEauthorblockN{Ahmet Yazar\IEEEauthorrefmark{1} and H\"{u}seyin Arslan\IEEEauthorrefmark{1}\IEEEauthorrefmark{2}}
\IEEEauthorblockA{\IEEEauthorrefmark{1}Department of Electrical and Electronics Engineering, Istanbul Medipol University, Istanbul, 34810 Turkey\\
}
\IEEEauthorblockA{\IEEEauthorrefmark{2}Department of Electrical Engineering, University of South Florida, Tampa, FL 33620 USA\\
Email: \{ayazar,huseyinarslan\}@medipol.edu.tr}
}

\maketitle

\begin{abstract}
Flexibility is one of the essential requirements in future cellular communications technologies. Providing customized communications solutions for each user and service type cannot be possible without the flexibility in 5G and beyond. Different optimizations need to be done for the flexibility related structures of 5G and beyond systems. In this paper, a novel machine learning (ML) based selection mechanism for the configurable waveform parameters is designed from the flexibility perspective. Moreover, a simulation based dataset generation methodology is proposed for ML systems. Results of computer simulations are presented using the generated dataset.
\end{abstract}

\acresetall


\IEEEpeerreviewmaketitle


\section{Introduction}
\label{sec:introduction}

5G systems come with various implementation flexibilities as a result of a large diversity of user requirements and service types along with wide variety of wireless channel conditions \cite{yazar2018a}. This flexibility is an important step for the future of communications systems. It is possible to see further flexibility trend in 6G radio networks \cite{saad2019a, raghavan2019a, tariq2019a, giordani2019a, david2018a}. New flexibilities inherently require new decision and selection mechanisms. Hence, there can be a need for different type of solutions together with the conventional ones while the flexibility is increasing.

\begin{figure}[bp!]
  \centering
  \includegraphics[width=7.0cm]{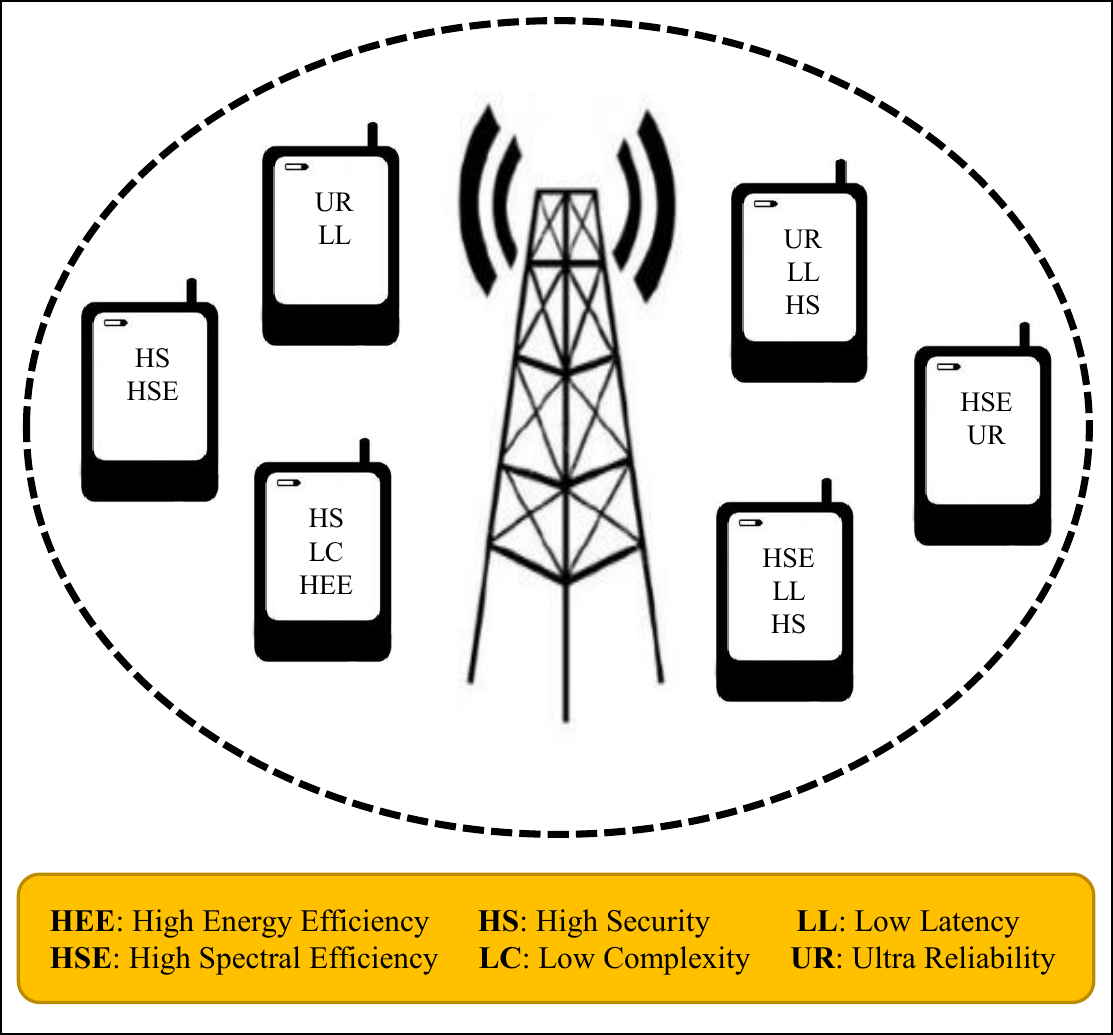}
  \caption{Example requirements of different service types.}
\label{fig:Fig1}
\end{figure}

\begin{figure}[t]
  \centering
  \includegraphics[width=7.0cm]{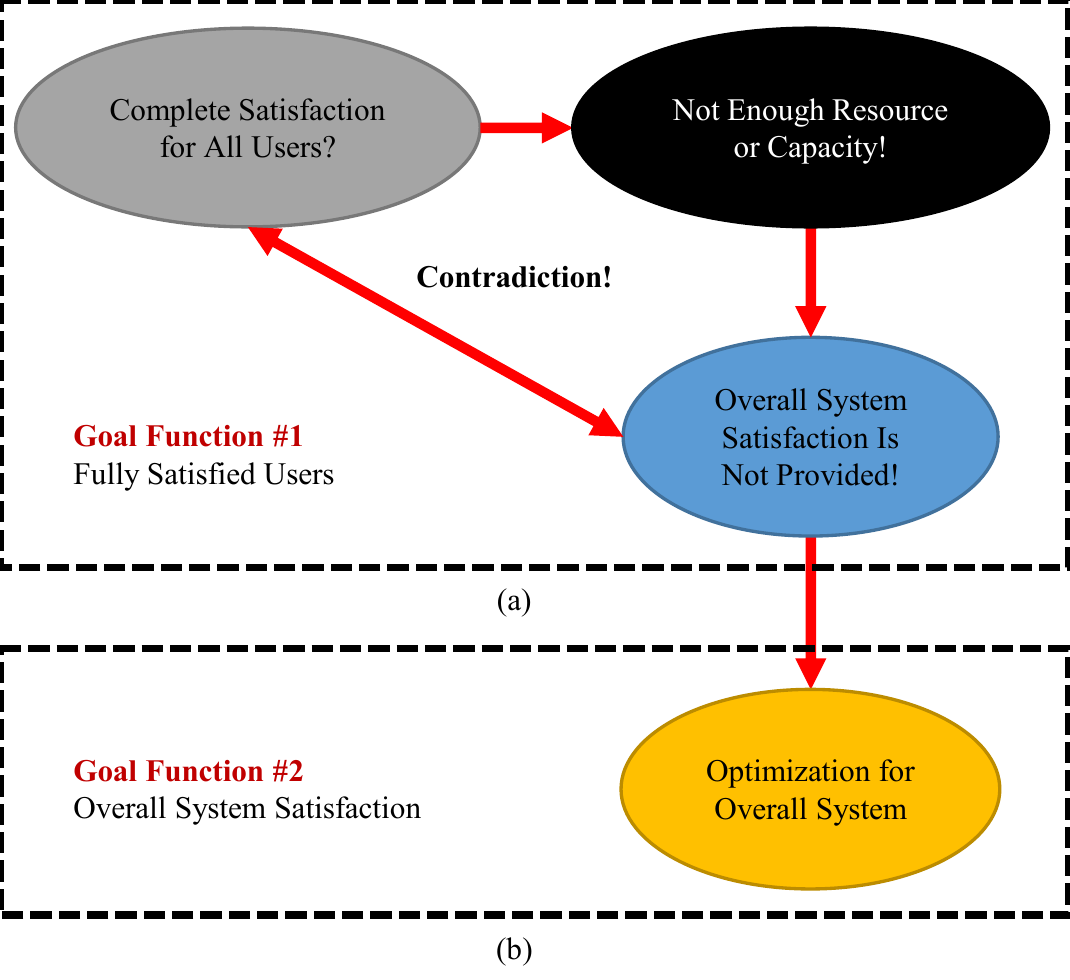}
  \caption{Relationship between the goal functions of a) fully satisfied users and b) overall system satisfaction.}
\label{fig:Fig2}
\end{figure}

From the flexibility perspective, ultra reliability, low latency, high security, high spectral efficiency, high energy efficiency, and low complexity are some example requirements of different service types as shown in Figure~\ref{fig:Fig1}. As a first basic problem, meeting some of these requirements together for a user is not an easy task under various constraints. Therefore, there is a need for proper optimization. The second basic problem is related to provide complete satisfaction for all users simultaneously as summarized in Figure~\ref{fig:Fig2}. It brings another optimization necessity because of the limited resources like in the knapsack problem. As can be seen from these basic problems, 5G and beyond flexibility comes with different optimization needs. There are several research opportunities for these optimizations in new generation communications systems through alternative solutions like Machine Learning (ML) that is one of the most popular recent trends not only for communications but also for various applications. In comparison with conventional methods, ML based solutions can be useful especially if there are many different relationships with the high number of parameters in a difficult problem.

ML can be employed for different optimization aims in wireless communications. Several ML paradigms and challenges for 5G systems are investigated in \cite{jiang2017a} and \cite{li2017a}. In \cite{mao2018a}, deep learning (DL) and wireless networks are discussed together in depth. Our paper tries to find an answer to the question of ``how 5G and beyond base stations decide on the waveform parameters of each user in a cell using ML''. There were 500, 1000, and 1500 parameters in 2G, 3G, and 4G, respectively \cite{imran2014a}. It is not difficult to except that 5G and beyond nodes will have more configurable parameters considering this trend. Related with these adjustable parameters, there is a large number of waveform parameters in 5G New Radio (NR). Most of these parameters are controlled by the base station. Conventional algorithms may not find the optimum parameters in such a rich set of options. The base station needs to provide the optimum waveform parameters for each user to solve two basic resource allocation problems that are given in Figure~\ref{fig:Fig2}. ML based methods can play an important role at this point. In the literature, end-to-end physical layer of wireless systems is analyzed from the DL perspective in \cite{oshea2017a}. In another study, authors develop an Orthogonal Frequency Division Multiplexing (OFDM) receiver and equalizer with ML \cite{zhao2018a}. These are some of the example ML studies that are related with resource allocation and waveform design for 5G and beyond.

A supervised ML based method is developed with this paper for the second goal function in Figure~\ref{fig:Fig2} to provide a general structure before the selection of waveform parameters for each user. For example, our method decides on the efficient number of numerologies (frame parameters) that can be assigned to users. However, it does not make user-numerology association directly. It can be done with conventional methods like in \cite{yazar2018a}. Also, different scheduling algorithms like \cite{yazar2018b} can be applied to enhance reliability and spectral efficiency  after the decision of general structure for configurable waveform parameters. In this paper, the main focus is on the general structure as a first step and it is done with ML methods. Contributions include designing a ML based selection mechanism for the configurable waveform parameters in 5G and beyond communications and introducing a simulation based dataset generation methodology for the objective of the waveform parameter selection.

The rest of the paper is organized as follows: Section~\ref{sec:assumptions} presents system model and assumptions. Waveform parameters and possible class labels for the dataset are described in Section~\ref{sec:parameters}. Feature extraction and automatic class labelling processes are explained in Section~\ref{sec:features}. In Section~\ref{sec:simulation}, simulation results of ML methods are provided. Finally, Section~\ref{sec:conclusion} gives the conclusion.

\section{System Model and Assumptions}
\label{sec:assumptions}

\begin{figure}[t]
  \centering
  \includegraphics[width=7.0cm]{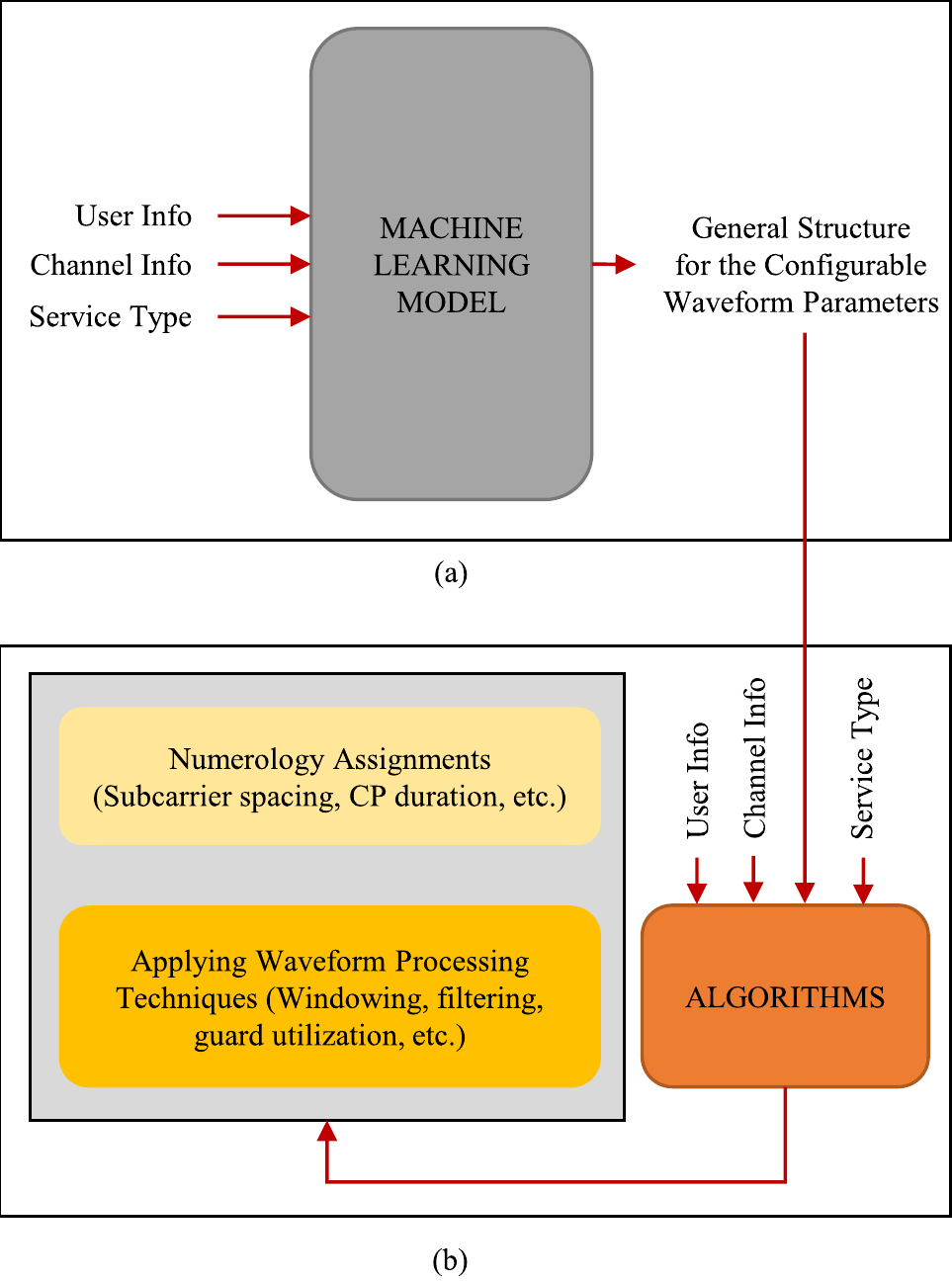}
  \caption{Example demonstrations for a) decision on general structure for the configurable waveform parameters and b) assignments of user-based waveform parameters.}
\label{fig:Fig3}
\end{figure}

The proposed supervised ML based method is designed to use in the first step of the overall system model as shown in Figure~\ref{fig:Fig3}(a). Decisions of the first step need to be employed while deciding on the user-based parameter assignments in BS. The first step of the model aims to increase overall system satisfaction.

The number of different service type element of the scenarios is three in 5G systems and they are listed as enhanced mobile broadband (eMBB), ultra reliable low latency communications (uRLLC), and massive machine type communications (mMTC). These service types and their requirements are used in the algorithms. Users can belong to one of 5G services and different Rayleigh fading channel models are used for each user. Therefore, the proposed ML system model can be assumed as channel-aware and service type-aware.

Dataset is generated using the results of three performance metrics in multi-numerology based 5G NR simulation. We assume that the 5G NR simulation gives perfect results. Performance metrics include signal to interference plus noise ratio (SINR), spectral efficiency, and flexibility. It is assumed that flexibility changes directly proportional to the number of numerologies. 5G NR numerology parameters are used according to 3GPP standard documents \cite{3gpp.38.211}. We allocate users with same numerologies contiguously in the frequency domain. It is assumed that each numerology block that consists of multiple carriers can be shared by multiple users.

\section{Waveform Parameters and Possible Class Labels for the Dataset}
\label{sec:parameters}

BSs make numerology assignment and resource allocation for each user according to user feedbacks. Additionally, BSs decide on the optional waveform processing techniques include windowing, filtering, guard utilization, and CP utilization. The difficulty of these decisions comes with the diversity of different requirements. Besides, the decisions need to be given considering all users together in a general structure as shown in Figure~\ref{fig:Fig3}. In this section, various waveform parameters and then possible class labels for a dataset are provided.

\subsection{Waveform Parameters in 5G and Beyond}

Long Term Evolution (LTE) employs Cyclic Prefix (CP) OFDM waveform with single numerology in downlink. CP-OFDM waveform with multiple numerologies is preferred in downlink of 5G NR. There are also optional waveform processing techniques like windowing and filtering in 5G NR. If it is assumed that there are $N$ options for multiple numerologies and $P$ options for waveform processing techniques, 5G NR has $NxP$ different options in the CP-OFDM based transmitter.

Multi-numerology structure of 5G NR provides different frame parameters for users in a cell. Four numerologies are defined for data transmission in 5G NR \cite{dahlman2018a}. In the future, more numerology structures can be defined to increase flexibility. Multi-numerology structures bring good flexibility in 5G but there is only one adjustable parameter which is a subcarrier spacing \cite{zaidi2018a}. The other numerology parameters depend on the subcarrier spacing CP duration, slot duration, maximum allowed bandwidth, and the number of slots per subframe are dependent to subcarrier spacing because of the simplicity and practicality. If it is assumed that users with same numerologies are located contiguously in the frequency domain and it is not mandatory to employ all of the numerologies, there can be $N=64$ numerology options constituted with four numerologies. If the number of waveform related independent adjustable parameters increases with beyond 5G, there are more options for multiple numerologies \cite{yazar2018a}. For example, adjustable CP duration and utilization is an important concept for beyond 5G. Additionally, using one common CP for different numerologies can also be another new concept in 5G beyond and it also changes the number of numerology options \cite{yazar2018c}.

Waveform processing techniques of 5G transmitters include windowing usage, filtering usage, and guard utilization. All of these techniques are defined as optional in 3GPP \cite{levanen2019a}. There can be more optional techniques in 5G beyond systems. Multiple numerologies and the other non-orthogonality sources increase the necessity of waveform processing techniques \cite{yazar2018c}. These techniques require various variable parameters. For example, several prototype filters in the literature including rectangular, raised-cosine, Gaussian and so on are provided in \cite{sahin2014a}. Different filters and the related coefficients increase the number of options for waveform processing techniques.

\subsection{Possible Class Labels for the Dataset}

As a summary, there are $N$ different type of numerology groups and $P$ waveform processing techniques. As mentioned, there are four types of numerologies but 64 numerology groups in 5G. For the waveform processing techniques, the same windowing, filtering and guard utilization can be employed for all users or different options can be preferred for each user. Guard bands can be used between numerologies in the frequency domain.

\begin{figure}[t]
  \centering
  \includegraphics[width=7.0cm]{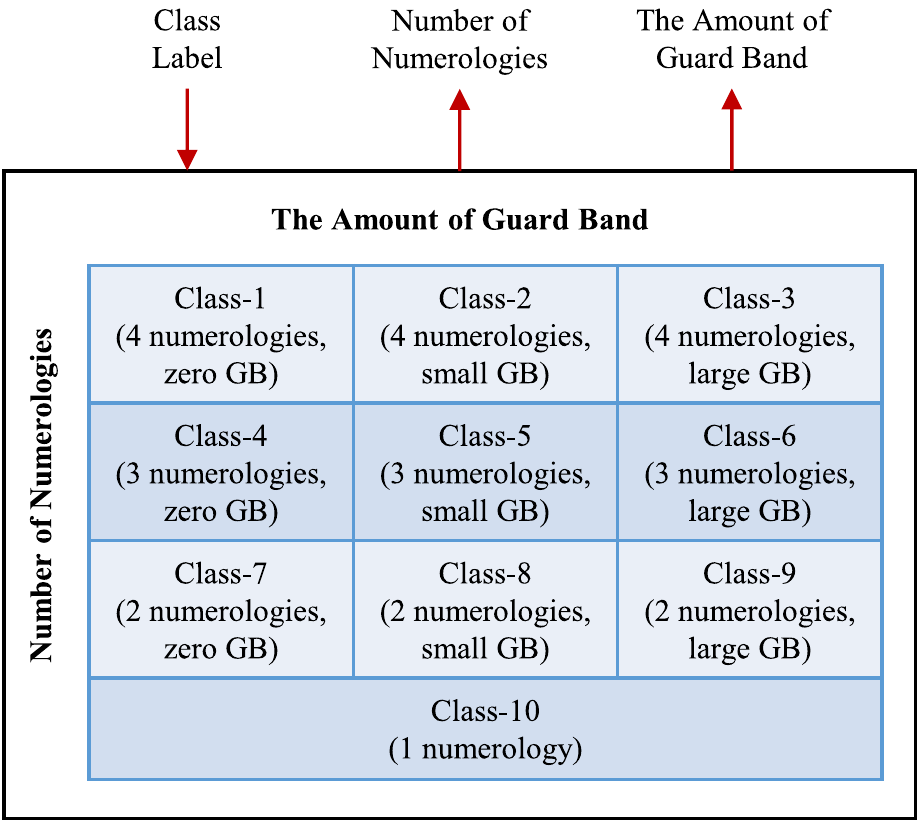}
  \caption{Details of the class labels that are included in the dataset.}
\label{fig:Fig4}
\end{figure}

There are $NxP$ classes under these conditions. Each class gives a different set of waveform parameters. The class labels can be defined with a multidimensional lookup table like in Figure~\ref{fig:Fig4}. More than 1000 classes can be considered for 5G cellular systems. The number of classes may increase exponentially for 5G beyond.

The number of class labels has direct effects on the difficulty of learning problem and computational complexity of training ML models. Additionally, if there are more class labels, a larger dataset is needed. Because of these restrictions, we only takes 10 classes in this paper for the sake of simplicity. It is assumed that there are four options about the numerology related parameters. Using four numerologies, three numerologies, two numerologies, and only one numerology are four basic different options. The other details of user-numerology association can be done in the next steps without ML. On the other side, it is assumed that there is only three guard band options as the waveform processing techniques. Windowing and filtering options are not included at this stage. The details of 10 class labels are given in Figure~\ref{fig:Fig4}. There is not any guard band necessity in Class-10 because there is only one numerology for this class.

\section{Feature Extraction and Automatic Class Labelling}
\label{sec:features}

\subsection{Feature Extraction}

It assumed that there are $U$ users in a cell. For all these users, independent random data is generated for $S$ different scenarios (e.g. thousands of random scenarios) based on the channel information in \cite{yarkan2008a}. Scenarios are defined with the parameters of random maximum excess delay, random maximum Doppler effect, and random service type (eMBB, uRLLC, or mMTC). If there are $I$ input random parameters (e.g. I is three in our example), $UxI$ parameters are generated for each scenario. Therefore, the generated raw dataset includes $S$ rows that are equal to the number of different data. The raw dataset also includes $UxI$ columns. In the future cellular systems, there can be more type of services. In addition, there will be more different type of new channel models for 5G beyond, especially for millimeter wave systems. It means that $S$ needs to be increased with the future wireless communications.

\begin{figure}[t]
  \centering
  \includegraphics[width=7.0cm]{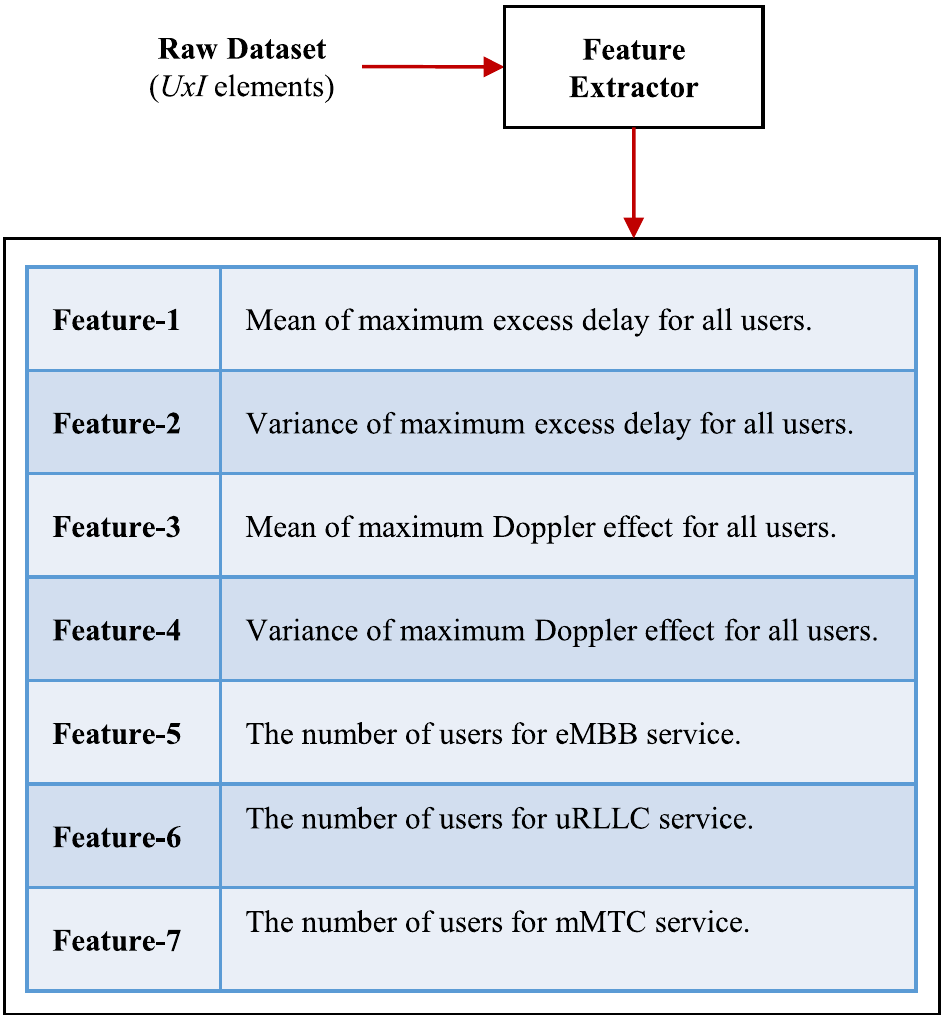}
  \caption{Details of the features that are included in the dataset.}
\label{fig:Fig5}
\end{figure}

Each data in the raw dataset includes $UxI$ parameters. $I$ is taken as three in our design. Feature extraction methods are chosen to make the design independent from the number of users. Seven features are extracted to make classification process efficiently. Details of features are provided in Figure~\ref{fig:Fig5}. The number of features can change.

\subsection{Simulation Based Automatic Class Labelling}

\begin{figure}[t]
  \centering
  \includegraphics[width=7.0cm]{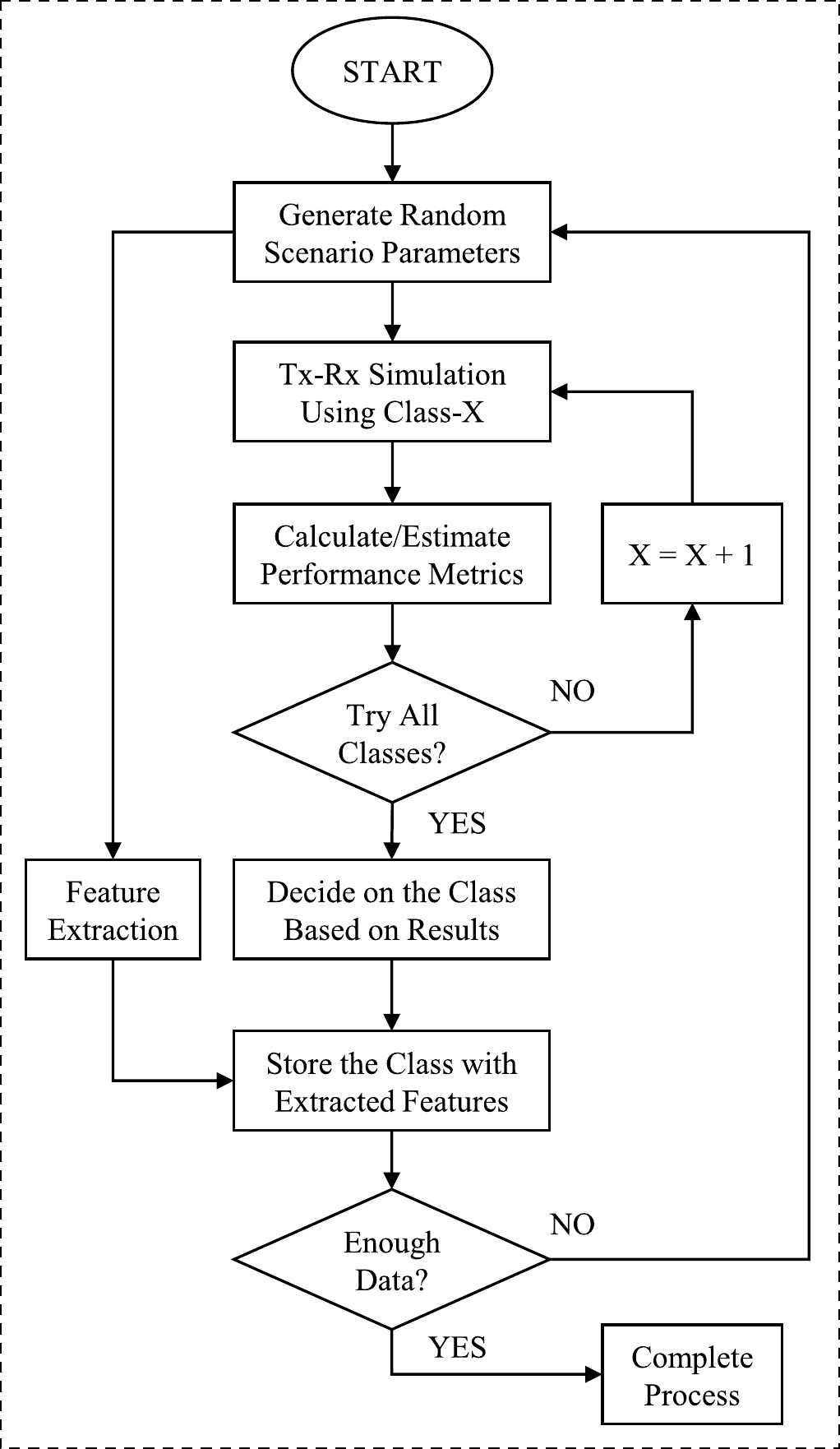}
  \caption{Algorithm flowchart for the simulation based dataset generation methodology. Automatic class labelling is done via simulation.}
\label{fig:Fig6}
\end{figure}

In simulation based dataset generation, there is a need to analyze the results of one or more performance metrics while deciding on the class labels automatically. The target performance metrics should be selected carefully for each scenario. All options of the waveform parameter sets are tried one by one on the simulation setup and three performance metrics (SINR, spectral efficiency, and flexibility) are calculated in each time. Then, the most successful waveform parameters are determined as a class label as shown in Figure~\ref{fig:Fig6}. Different amount of priorities and weights to the related performance metrics are given considering the service type majority. As an example, spectral efficiency metric has more priority for eMBB service but SINR metric has more priority for uRLLC service. If there are high number of users with all type of services, then the flexibility metric has also priority because the overall system needs to meet with many different requirements together.

After the automatic class labelling process, the number of samples in each class is counted. If there is a high variance for the number of samples in each class, some of the samples in the most frequent classes are removed from the dataset. Otherwise, ML training can be biased for the most frequent classes in the dataset.

\section{Simulation Results}
\label{sec:simulation}

In the dataset, there are 114420 samples (rows) with seven feature columns and one class label column corresponding to random scenarios. Class distributions are equal. The dataset is divided as training, validation, and testing with 80\%, 20\%, and 20\% ratios, respectively. Automatic class labelling is employed using a multi-numerology based CP-OFDM simulator that is designed for 3GPP 5G NR standards.

\begin{figure*}[t]
  \centering
  \subfigure[Confusion matrix for KNN classifier with five neighbors.]{\includegraphics[width=6.0cm]{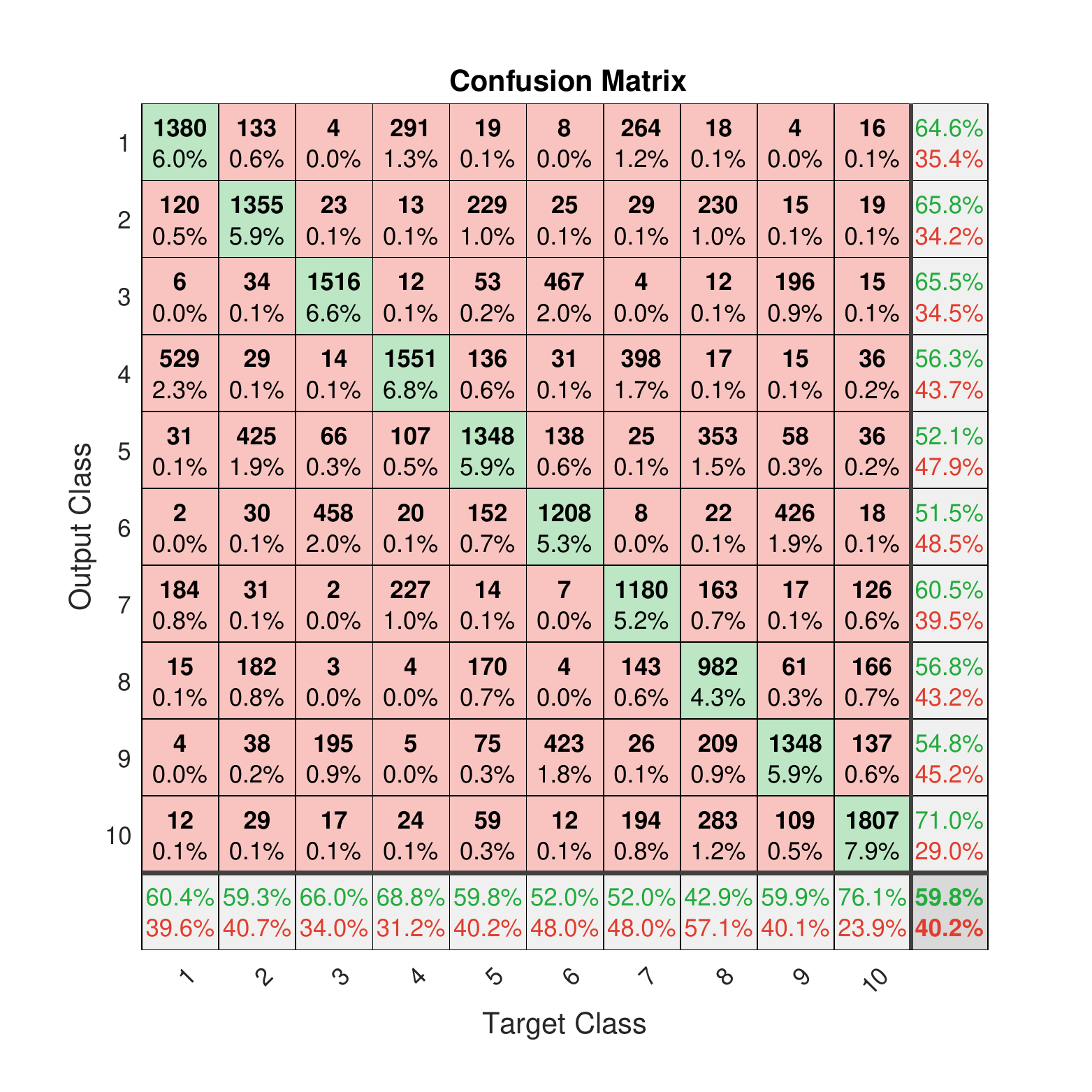}}\qquad
  \subfigure[Confusion matrix for decision tree classifier.]{\includegraphics[width=6.0cm]{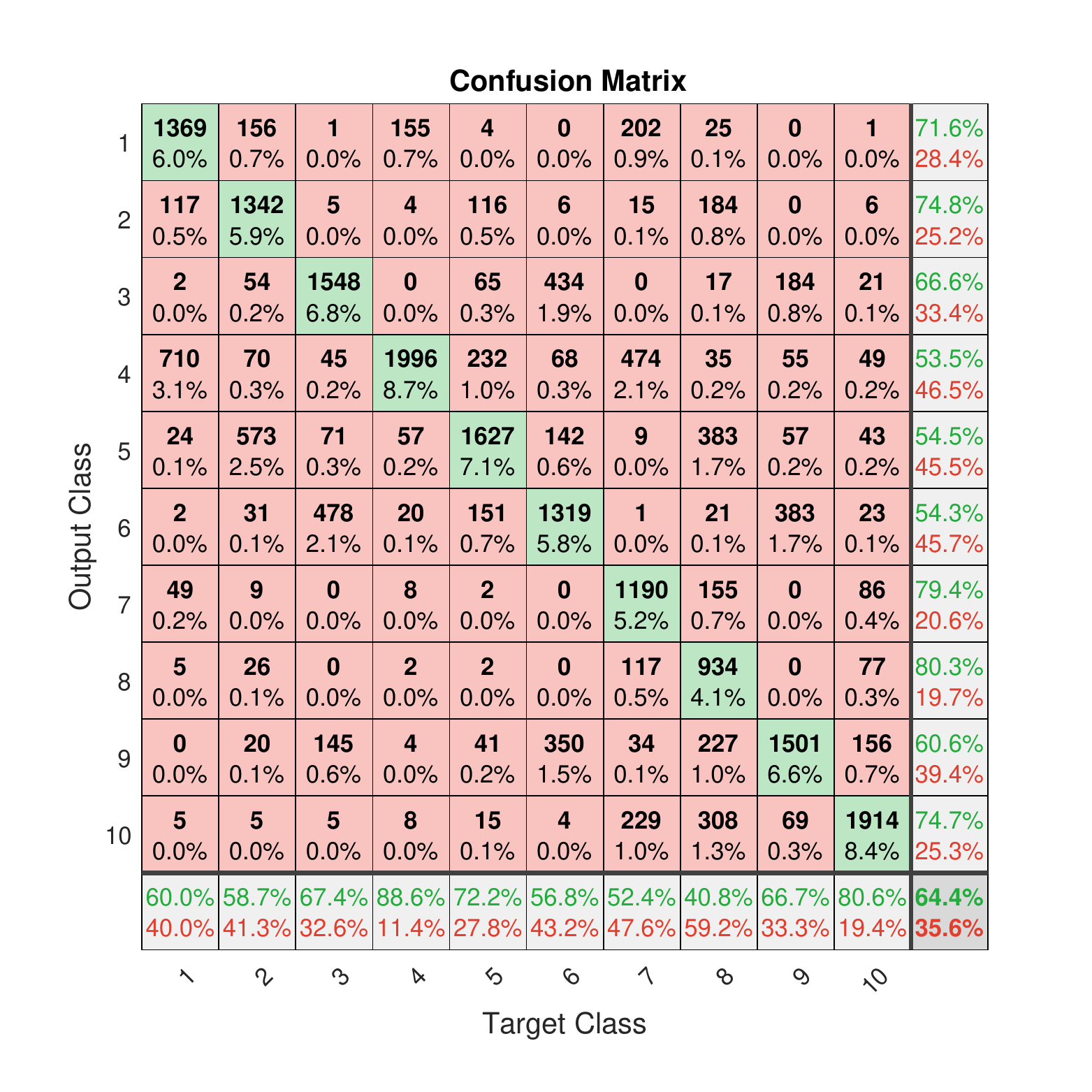}}\\
  \subfigure[Confusion matrix for NN networks with 20 hidden neurons using scaled conjugate gradient backpropagation algorithm.]{\includegraphics[width=6.0cm]{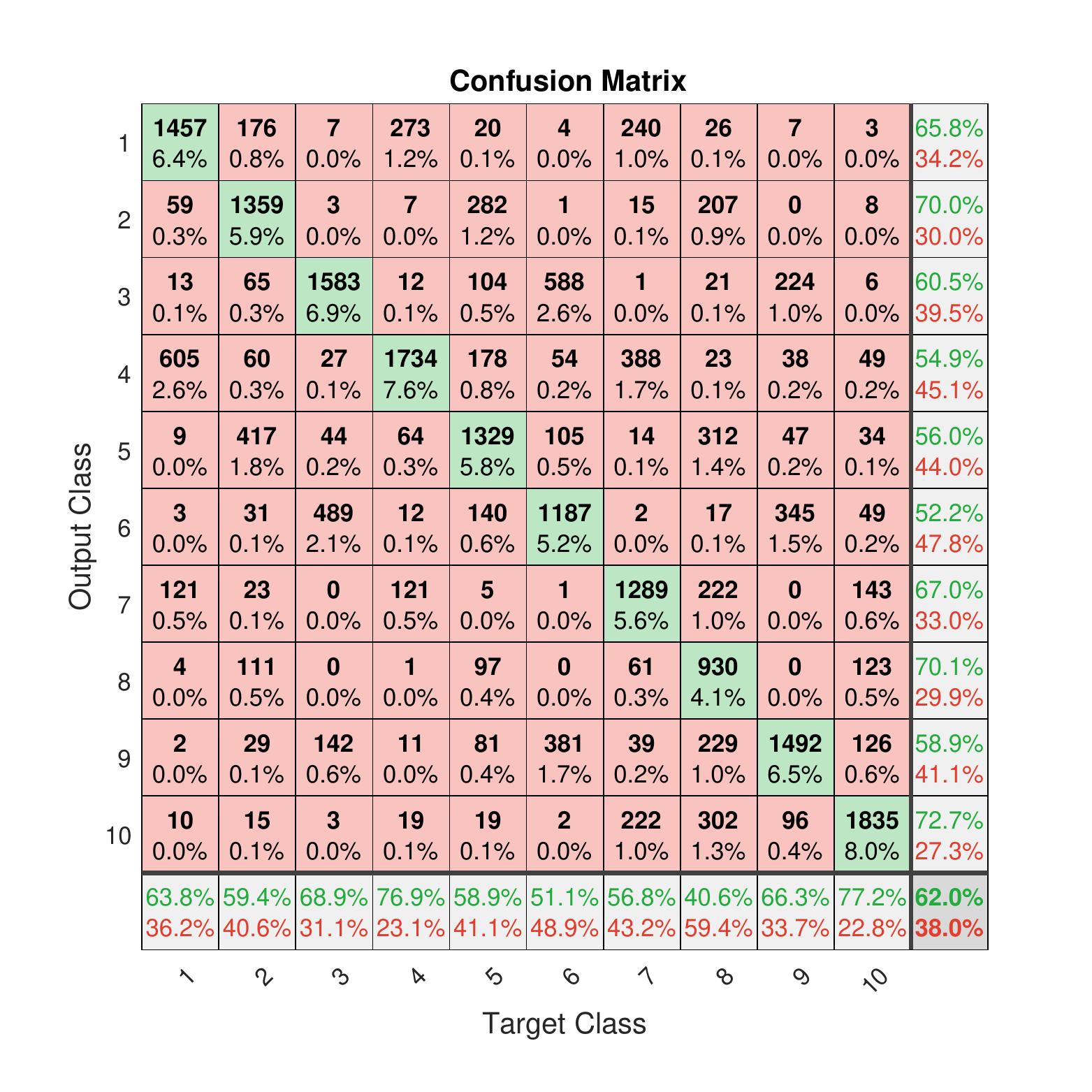}}\qquad
  \subfigure[Confusion matrix for NN networks with 20 hidden neurons using Bayesian regularization backpropagation algorithm.]{\includegraphics[width=6.0cm]{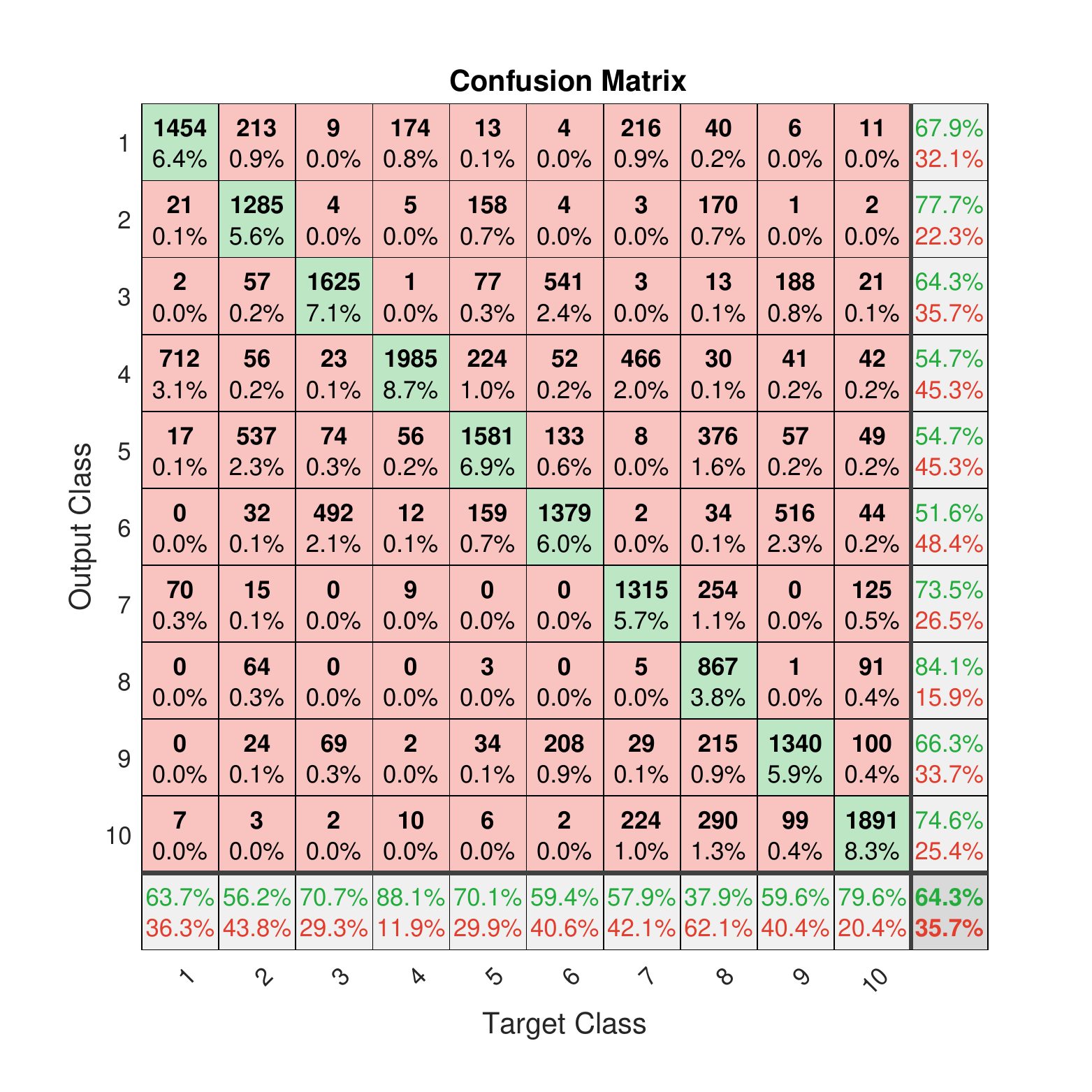}}\\
  \subfigure[ROC curve for NN networks with 20 hidden neurons using Bayesian regularization backpropagation algorithm.]{\includegraphics[width=6.0cm]{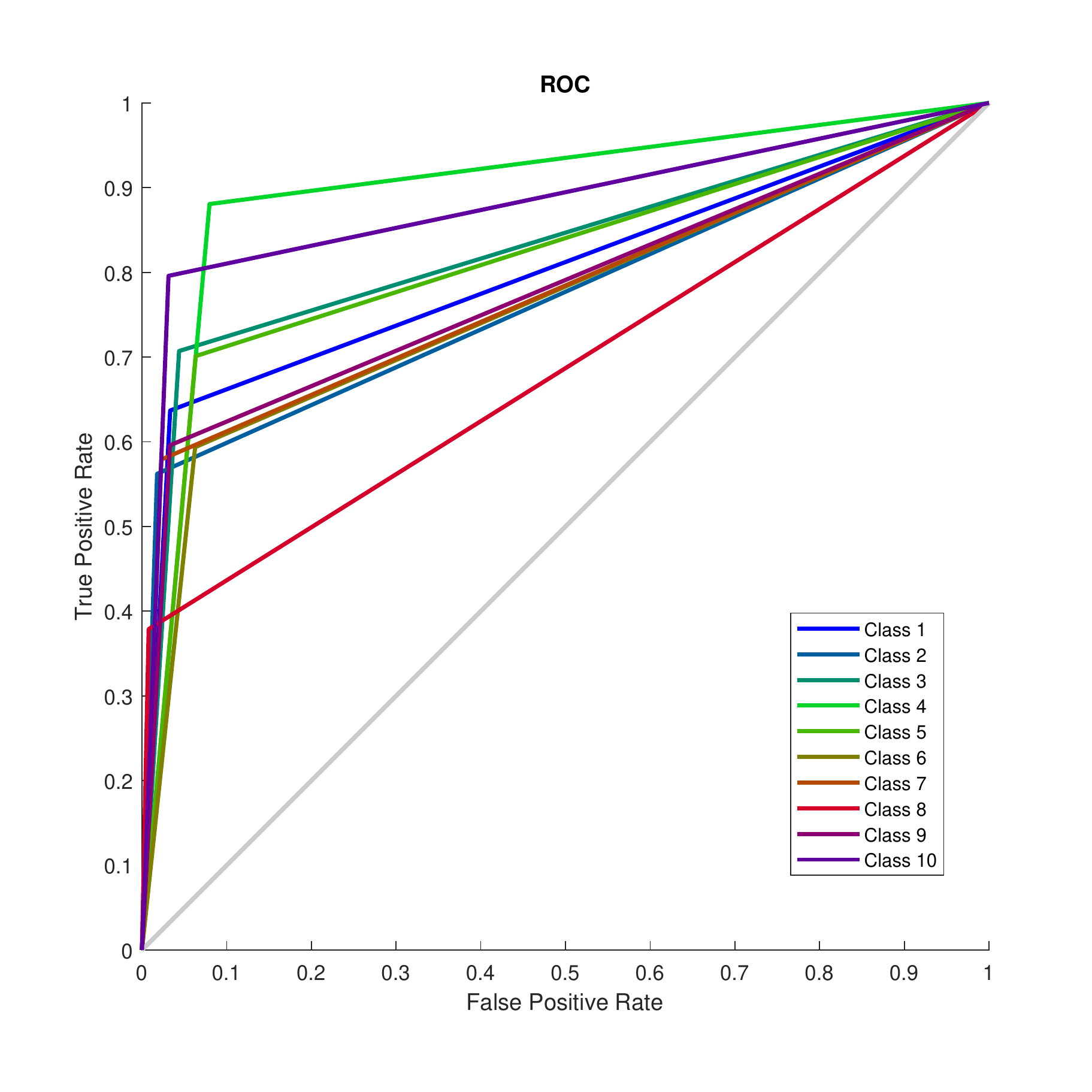}}\qquad
  \subfigure[ROC curve for NN networks with 20 hidden neurons using Bayesian regularization backpropagation algorithm. Neighbour class labels are grouped.]{\includegraphics[width=6.0cm]{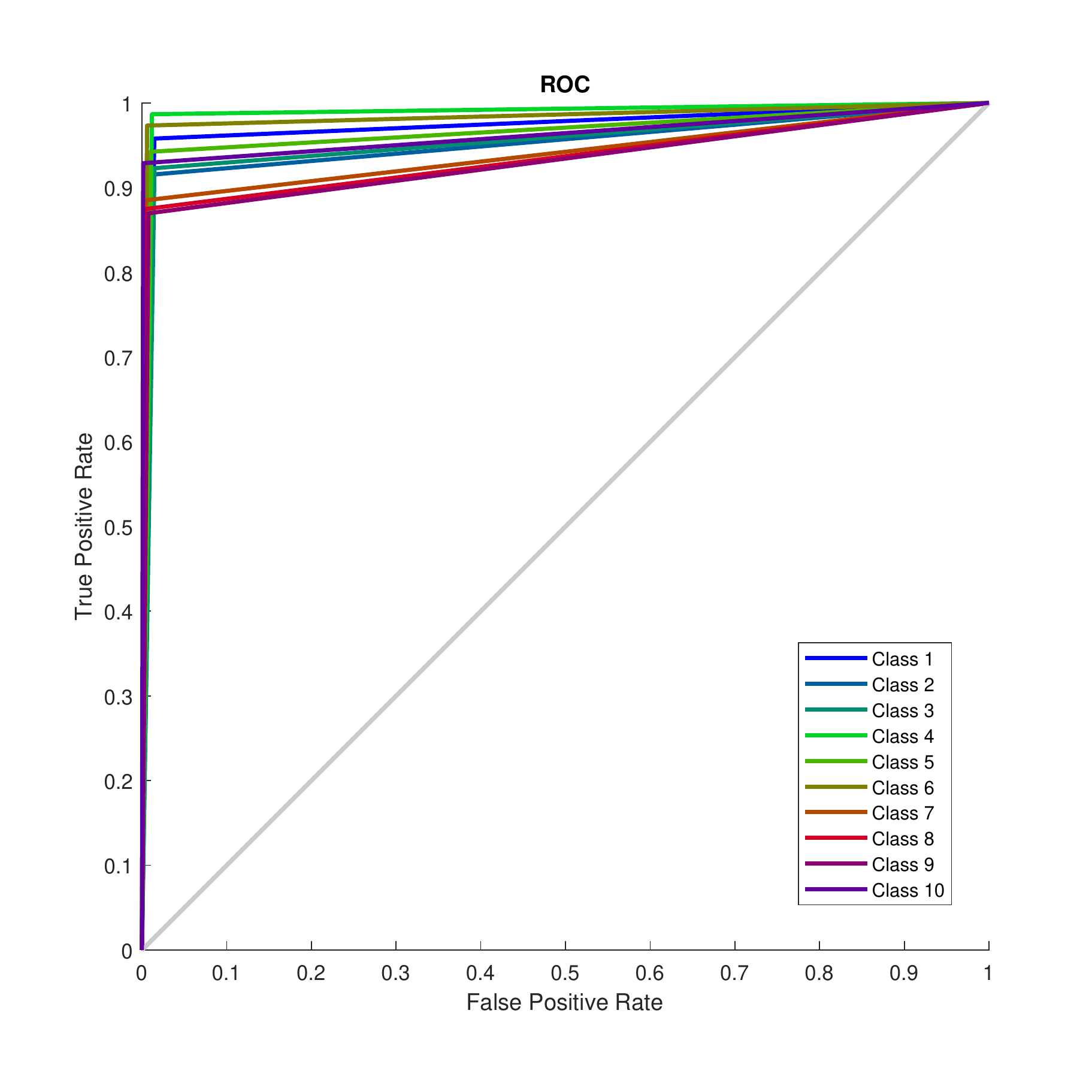}}\\
  \caption{Confusion matrices and ROC curves for the simulation results in MATLAB platform.}
\label{fig:Fig7}
\end{figure*}

MATLAB platform is preferred in the computer simulations. 'fitcecoc' and 'patternnet' functions of Statistics and Machine Learning Toolbox and Deep Learning Toolbox, respectively, are used while training various ML models. Hyperparameter optimizations are done with the same functions. Discriminant classifier, KNN clasifier, decision tree clasifier, naive Bayes classifier, and neural networks (NN) are employed during the simulations. The results of relatively more succesful models are presented in Figure~\ref{fig:Fig7}. Confusion matrices show that success rates changes between 60\% and 65\% for 10 classes. Additionally, if neighbour classes in Figure~\ref{fig:Fig4} are grouped together, the success rates vary between 90\% and 93\% for the same classifier models. For example, if the decision for number of numerologies three or four, it can be acceptable while neighbour classes are grouped. However, confusion matrices in Figure~\ref{fig:Fig7} are provided without grouping neighbour classes. Receiver operating characteristic (ROC) curves in Figure~\ref{fig:Fig7}(a) and Figure~\ref{fig:Fig7}(b) shows the results without and with grouping neighbour classes, respectively.

As it can be seen from the simulation results, selection of waveform parameters is not an easy task. If the number of class labels (parameter sets) is increased, the problem will be more difficult. Modulation types and orders, the other waveform processing techniques like windowing and filtering parameters can be included for the class labels. In this case, there are more than 1000 class labels in a dataset. Hence, ML mechanisms may not be sufficient by itself under these conditions. Conventional methods need to be employed with ML mechanisms together to obtain better success results.

\section{Conclusion}
\label{sec:conclusion}

5G and beyond BSs need ML based decision units and optimization modules. This paper provides a detailed example to understand how can we use ML in 5G and beyond wireless systems. The proposed dataset generation methodology can be used to develop large datasets for better ML models in several studies. As a future work, the remaining waveform processing techniques like windowing and filtering will be included in our class labelling system. Also, different modulation types can be included with bit error rate (BER) performance metric. DL based testbed will be tested with a larger dataset. Moreover, the other steps like user-numerology association will be carried out with ML. As a concluding remark, feasibility of ML based systems should be analyzed to see that ML provides a better solution compared to the practical and effective conventional solutions.


\end{document}